\documentclass{article}
\usepackage[utf8]{inputenc}
\usepackage{amsthm}
\usepackage{graphicx,ae,aecompl}
\usepackage{amssymb}
\usepackage{amsmath}
\usepackage{subfigure}
\usepackage{algorithmic}
\usepackage[boxruled,linesnumbered]{algorithm2e}
\usepackage[left=2.5cm,top=2.5cm,right=2.5cm,bottom=2.5cm]{geometry}
\usepackage{makeidx}
\usepackage{float}
\usepackage{scalefnt}
\usepackage{fancyhdr}

\usepackage{lipsum}
\usepackage{xcolor}
\usepackage{hyperref}
\usepackage{authblk}

\usepackage{booktabs}

\title
{Model selection criteria for  regression models with splines and the automatic localization of knots} 

\author{Alex R. dos S. Sousa, Magno T.F. Severino and Florencia Leonardi
\\ Institute of Mathematics and Statistics, University of São Paulo, Brazil}

\begin{document}

\numberwithin{equation}{section}
\numberwithin{table}{section}
\numberwithin{figure}{section}

 \maketitle
  \begin{abstract}
This paper proposes a model selection approach to fit a regression model using splines with a variable number of knots.  We introduce a penalized criterion to estimate the number and the localization of the knots where to anchor the spline bases.  The method is evaluated on simulated data and applied to Covid-19 daily reported cases for short-term prediction.  
\end{abstract}

\section{Introduction}
Spline regression is one of the most useful tools for
nonparametric and semiparametric regression and has been extensively developed along the last decades. 
The model works by selecting some knots on the function domain to define a partition 
and then fit a fixed degree polynomial on each segment in such a way that these 
polynomials smoothly match at the internal knots through continuity conditions imposed 
on their derivatives. 
The result is typically a smooth curve fitting, which accurately estimates a large variety of smooth 
functions. Further, spline regression offers a great flexibility in curve estimation problems according 
to convenient choices of degree and internal knots and, as generally occurs in nonparametric 
regression by basis expansion, the infinite estimation problem becomes a finite parametric 
estimation problem, since the spline function is determined by the estimation of its coefficients. For 
more details about spline based methods, their theory and applications in Statistics, see 
Wahba (1990), Ramsay (2004), Friedman et al. (2001), and James et al. (2013). 

One of the main issues in spline regression is the selection of the number and positions of the
knots. In some real data analyses, wrong specification of the knots can cause a high impact 
on the spline regression performance. 
Excessive number of knots may lead to overfitting while an
insufficient number of knots can lead to underfitting. Moreover, the positions of the knots should be 
well determined, since high concentration of knots on a specific interval of the domain overfits 
the data on this region and underfits the data outside it. 
Thus, a good procedure for knots selection should provide optimal number and location of  
knots, in order to have a well performed spline curve fitting. 

In practice, the knots are manually selected 
and few studies in the
literature provide some systematic method to overcome this difficulty. Friedman and Silverman (1989), 
 Friedman (1991) and Stone et al. (1997) studied stepwise based 
 methods for a set of candidates of knots. Osborne et. al (1998) proposed an algorithm 
 based on the LASSO estimator to estimate the locations of knots. Ulker and Arslan (2009) proposed an artificial immune system to perform knots selection in curve
  and surface B-splines estimation by considering the knots locations candidates as antibodies. 
  Bayesian methods are proposed by Denilson et al. (1998) who proposed a joint prior 
  distribution for the number and locations of knots, and Biller (2000), who gave an 
  adaptive bayesian approach for knot selection based on reversible jump Markov chain Monte Carlo
   in semiparametric generalized linear models.
   
    In penalized spline regression, Eilers and Marx (1996), who defined the term P-splines, generalized the ideas of O'Sullivan (1986) by proposing the use of order differences of the spline coefficients as penalization criteria and the knots were set equidistantly.  Ruppert (2002) 
 presented two algorithms for selection of a fixed number of knots on penalized spline regression. The first one, called myopic algorithm, is based on the minimization of the generalized cross validation statistic and the second one, the full search method, is based on the Demmler-Reinsch type diagonalization. Later, Ruppert et al. (2003) proposed a default choice of the number of knots as the minimum value between one quarter of the number of different domain observed values and 35. Both Ruppert (2002) and Ruppert et al. (2003) set the knots at the quantiles of the data, which can lead to a lack of knots in regions of the function that contain important features such as peaks and local maximum and minimum, for example. 
To address this problem, Yao and Lee (2008) proposed a complement in setting knots at quantiles of data by adding more knots in local characteristics of the function.  Kauermann and Opsomer (2011) considered the number of knots in penalized spline regression as a parameter to be estimated by likelihood. Again, as the previous works, the locations of knots are the quantiles of data. An automatic selection procedure to determine locations of knots was proposed recently in an arXiv preprint paper by Goepp et al. (2018) and was called Adaptive Spline or A-spline. The method essentially starts with a large number of knots and removes excessive knots by adaptive ridge algorithm based on adaptive penalized likelihood. Moreover, theoretical and asymptotic behaviour of the number of knots were studied in Kauermann et al. (2009), Claeskens et al. (2009) and Li and Ruppert (2008). Theoretical results about penalized spline regression can be seen in Hall and Opsomer (2005).      

Although the methods described above have been well succeeded, they work with a fixed number 
of knots  and/or require some prior information regarding possible regions where the knots 
should be set, as the quantiles of data for example, which can be seen as a limitation in some cases.    
In this work we propose an automatic method to estimate  both the number of knots and 
their locations,  based on the minimization of  a penalized  least squares criterion. 
The penalty plays the role of avoiding a high number of knots which consequently leads to overfit the data. 

We study the performance of the method on two simulated scenarios  to evaluate its ability of estimating the number of knots and their location.
Also, we compare the results with other well established methods.
Further, we also apply the procedure to daily reported cases of Covid-19 
in several countries to fit their epidemiological curve of cases and perform short term forecast.
Both simulation studies and real data application suggest that the proposed method can be considered by practitioners for application in spline regression problems.

This paper is organized as follows: Section~2 describes the spline model and the proposed method of automatic knots selection. Section~3 presents simulation studies regarding the performance of the method, and Section~4 show applications to real datasets related to daily reported cases of Covid-19. 
Finally, Section~5 provides conclusion and suggestions of future works.  

\section{Spline regression model with variable number of knots}

We start with a nonparametric regression problem of the form
\begin{equation*} 
y_i = f(x_i) + \epsilon_i , \qquad i=1,...,n,
\end{equation*}
\noindent where $x_i$ are scalars, $f$ is an unknown smooth function with domain $[a,b]$ such that $f(x_i) = \mathbb{E}(y_i|x_i)$ and $\{\epsilon_i\}_{i=1}^{n}$ are zero mean independent random variables with unknown variance $\sigma^2$. To estimate $f$, we define a sequence of knots $a=t_0 < t_1 <\cdots<t_{K}<t_{K+1}=b$ that defines a partition of $[a,b]$ in intervals $[a,t_1),[t_1,t_2),\ldots,[t_K,b]$ and use a $m-$th order spline regression model, $m=p+1$, defined by 
\begin{equation} \label{eq:splines}
f(x) = \beta_{0} + \sum_{j=1}^{p} \beta_{j}x^{j} + \sum_{k=1}^{K} \beta_{p+k}(x-t_{k})_{+}^{p},
\end{equation}
where $p, K \in \mathbb{Z}_{+}$, $\boldsymbol{\beta} = [\beta_0,\ldots,\beta_{p+K}]'$ 
is the vector of coefficients, $(u)_{+}^{p} = u^{p}\mathbb{I}(u \geq 0)$ and $t_1 <\cdots<t_{K}$ 
are fixed internal knots which, from now on,  will be referred to as just knots. 
In general, a spline regression model is a $p$ degree  polynomial on each interval of two 
consecutive knots and has $p-1$ continuous derivatives everywhere 
(supposing that the knots have no multiplicity). 
In this work we adopt 
a set of knots sufficiently spaced, i.e, we assume  $|t_{k+1} - t_{k}| > \delta$ for some 
$\delta >0$, $k=0,\ldots,K$. Thus, the problem of estimating $f$ becomes a finite 
parametric estimation problem, whose parameters are the vector of 
coefficients $\boldsymbol{\beta}$, the number of internal knots $K$ and their locations
$t_k$, $k=1,\ldots,K$. 
Let us denote by $\boldsymbol{\theta}$ the vector of parameters to be estimated, i.e, 
$\boldsymbol{\theta} = [\boldsymbol{\beta}, K, t_1,\cdots,t_K]'$. 
We assume a fixed value for $p$, the degree of the spline function. 
 

The estimation of the parameters takes into account the performance of the model regarding a loss function; in our case  the residual sum of squares. 
To avoid overffiting by choosing an excessive number of knots, we add 
a roughness penalty defined in terms of the 
number of intervals established by the knots, i.e, the $K$ knots define $K+1$ intervals 
on the domain of $f$. Thus, we define the penalized sum of squares $PSS_{\lambda}$ as
\begin{equation}\label{main-eq}
PSS_{\lambda}(\boldsymbol{\theta};\boldsymbol{y}) = \sum_{i=1}^{n}[y_i - f(x_i)]^2 + \lambda (K+1),
\end{equation}
where $\boldsymbol{y} = [y_1,...,y_n]'$, $\lambda>0$ is a tuning parameter, 
 and $f$ is the spline function in \eqref{eq:splines}. 
Thus, the estimator  $\boldsymbol{\hat{\theta}}$ is  obtained by
\begin{equation} \label{min-eq}
\boldsymbol{\hat{\theta}} = \arg \min_{\boldsymbol{\theta}} PSS_{\lambda}(\boldsymbol{\theta};\boldsymbol{y}).
\end{equation}

Considering the definition above, we search for a curve estimate that has good fitting properties with low residual sum of squares and which is also smooth enough according to a reasonable number of knots  
 that is controlled by the penalty on the number of segments of the domain.
  
The tuning parameter $\lambda$  attributes a weight on the roughness penalty term.
High values of $\lambda$ increase the importance of the penalization, 
that is, preference for smoothness over accuracy. 
On the other hand, low values of $\lambda$ favor models with high number of parameters, which can lead to overfitting. 
The specification of a satisfactory value for  $\lambda$ is an issue
 inherent to any regularized estimator and, as in other similar approaches, can be chosen by a standard cross validation procedure. 
  
Algorithm~1 describes the procedure to obtain $\boldsymbol{\hat{\theta}}$ in \eqref{min-eq}.
In fact, the algorithm requires as inputs the minimum distance $\delta>0$ between consecutive knots and the maximum number of knots $K_{max} \in \mathbb{N}$. 
For each $k \leq K_{max}$, the algorithm obtain the best vector of knots locations $(t_1,\ldots, t_k)$ in terms of ordinary least squares (OLS). 
Then the overall optimal vector is the one with the smallest $PSS_{\lambda}(\cdot;\boldsymbol{y})$. 
We developed the \textsf{splineSelection}\footnote{The \textsf{splineSelection} package is under development and is available on \textsf{Github}. To install it in \textsf{R}, run the command: \textsf{devtools::install.github(Alexestat/splineSelection)}.} \textsf{R} package to implement Algorithm 1, see Sousa et al. (2020) for details.
 
\begin{algorithm}[H]

 \textbf{Input:} $\lambda$, $\delta$, $K_{max}$ \\
  \For{$k \in \{1,...,K_{max}\}$}
    {
        Obtain $A_k = \{\boldsymbol{t}=(t^{(1)},\cdots,t^{(k)}): |t^{(i+1)}-t^{(i)}|>\delta \}$
        
        \For{$ \boldsymbol{t^{\star}} \in A_k$}
        { Obtain $\boldsymbol{\beta^{\star}}$ by OLS
        
          $\boldsymbol{\theta^{\star}} = \left(\boldsymbol{\beta^{\star}},k, t^{\star} \right)$ 
          
          Calculate $PSS_{\lambda,\delta}(\boldsymbol{\theta^{\star}};\boldsymbol{y})$

        }
        
    }
 \textbf{Output:} $\boldsymbol{\hat{\theta}} = \arg \min_{\boldsymbol{\theta^{\star},k}} PSS_{\lambda,\delta}(\boldsymbol{\theta^{\star}};\boldsymbol{y})$

\caption{Algorithm to obtain the proposed penalized least squares estimate of $\boldsymbol{\theta}$.}
\end{algorithm}
\subsection{B-splines and natural splines models}

\begin{figure}[H]
\centering
\includegraphics[scale=0.60]{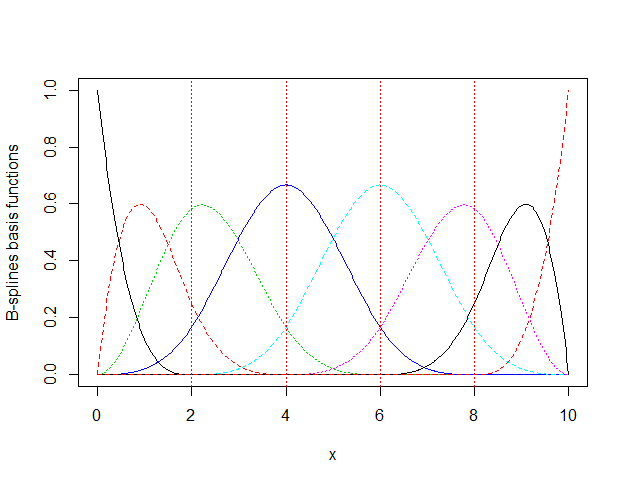}
\caption{Eight cubic B-splines basis defined by four interior knots.}\label{fig:bsplines}
\end{figure}

Our proposed method for automatic knots selection can also be applied to other spline basis structures. 
A well known spline basis is the B-splines, proposed by De Boor et al.(1978), which are more adapted for computational implementation due to its compact support. The $i$-th B-spline of order $m$ and knots $a=t_0 < t_1 <\cdots<t_{K}<t_{K+1}=b$, denoted by $B_{i,m}(x)$ is a spline of order $m$ defined on the same knots such that is nonzero over at most $m$ consecutive subintervals, i.e, over $[t_i,t_{i+m}]$ and zero outside it and can be defined recursively by
\begin{equation}
B_{i,m}(x) = \frac{x - t_i}{t_{i+m} - t_i}B_{i,m-1}(x) + \frac{t_{i+m+1} - x}{t_{i+m+1}-t_{i+1}}B_{i+1,m-1}(x),
\end{equation}    
where $B_{i,1}(x) = \mathbb{I}_{[t_i,t_{i+1})}(x)$. Then, $f$ is expanded in the B-splines basis instead of \eqref{eq:splines} as
\begin{equation}
f(x) = \sum_{i=1}^{m+K+1}\alpha_iB_{i,m}(x), 
\end{equation}
where $\boldsymbol{\alpha}=[\alpha_1,\ldots,\alpha_{m+K+1}]'$ is the vector of the coefficients of the B-splines expansion. 
Note that, in this setup, we have that $\boldsymbol{\theta} = [\boldsymbol{\alpha}, K, t_1,\ldots,t_K]'$.
Figure \ref{fig:bsplines} presents eight cubic B-splines (B-splines of order 4) basis functions defined by four interior knots on the interval $[0,10]$.
 
Another commonly used spline basis structure is the natural spline. This approach imposes 
the function $f$ to be linear on its extremes, i.e,
in regions lesser than $t_1$ and greater than $t_K$. 
Such constraint guarantees stable estimation on the boundaries of the function, 
see James et al. (2013). 
For more about spline basis structures, see De Boor (2001).

\begin{figure}[H]
\centering
\subfigure[]{
\includegraphics[width=8cm]{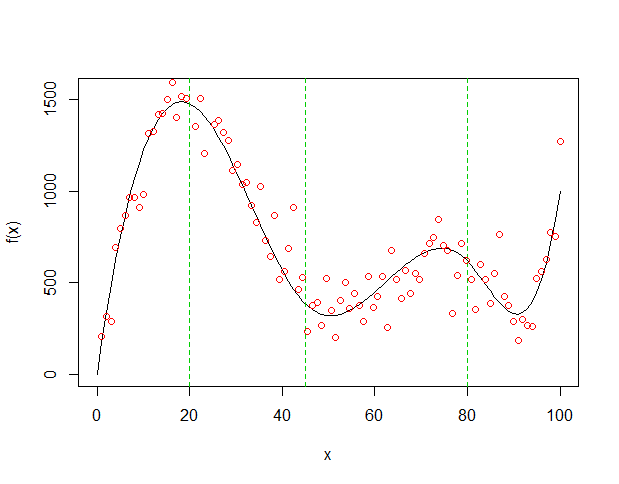}}
\subfigure[]{
\includegraphics[width=8cm]{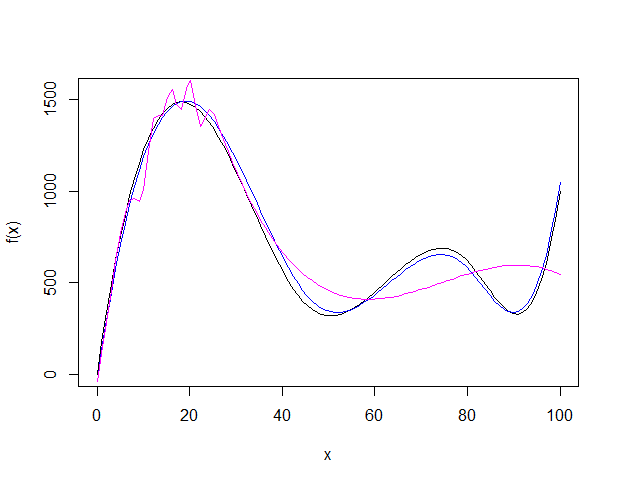}}
\caption{(a) Underlying function (black line), data points generated by this function. and knots locations (green vertical dashed lines). (b) Underlying function (black), estimated curve by cubic B-splines with 11 knots at $[t_1, t_2,\cdots,t_{11}]'=[6, 8, 10, 12, 14, 16, 18, 20, 22, 24, 26]'$ (pink), 
and estimated curve by the proposed method (blue) with optimal number and locations of knots.}\label{fig:example1}
\end{figure}

To illustrate the role of the penalizing term
 $\lambda$ in \eqref{main-eq},
we generate data from an underlying function defined by cubic B-splines basis on $[0,100]$ 
with three not equally spaced knots at $t_1 = 20$, $t_2=45$, and $t_3=80$ 
and signal to noise ratio SNR=3. Figure \ref{fig:example1}(a) shows the underlying function, its knots locations, and the generated data. 
After data generation, we estimate the underlying function using 
cubic B-splines expansion considering eleven knots 
fixed at
$[t_1, t_2,\ldots,t_{11}]'=[6, 8, 10, 12, 14, 16, 18, 20, 22, 24, 26]'$. 
The estimated curve is the pink line in Figure \ref{fig:example1}(b). 
Note that we 
intentionally set several knots at one specific region of domain, around the 
maximum
of the curve, and this inappropriate choice of number and location of knots may lead to two possible issues.
First, the estimated curve overfits data at the region 
where there are the most number of knots, that is,
the curve fit did not recover the peak with the required 
degree of smoothness, yielding some kind of interpolation 
in this region. 
Second, 
the absence of knots 
in the remaining regions did not allow the correct estimation of the
underlying function 
leading to underfit of the data in these regions.
For this reason, it is important to consider a penalty term that takes 
account this trade-off between number/locations of knots and smoothness. 
Penalizing the least 
squares by the number of segments determined by the knots partition guarantees the estimated 
curve to have this optimal number/locations of knots to avoid both under and overfitting problems. 
Moreover, this procedure is automatically performed, thus requiring no
previous knowledge regarding the number and positions of knots, which is of great interest and an advantage of our method. The blue line of Figure \ref{fig:example1}(b) is 
the estimated curve provided by the proposed method. The number of knots was correctly 
determined, $\hat{K}=3$ and the estimated knots locations were $\hat{t}_1=21$, $\hat{t}_2=47$ and 
$\hat{t}_3=79$.        

\section{Simulation studies}
In this section we present two simulation studies to evaluate the performance of the proposed method.
Section~\ref{sec:sim1} aims to analyse the method's ability to correctly estimate the number and locations of the knots for data generated by cubic B-splines and Section~\ref{sec:sim2} compares the proposed method's performance with two penalized spline regression methods. 

\subsection{Simulation study 1} \label{sec:sim1}
\begin{figure}[H]
\centering
\subfigure[$t_1 = 50$.]{
\includegraphics[width=5cm]{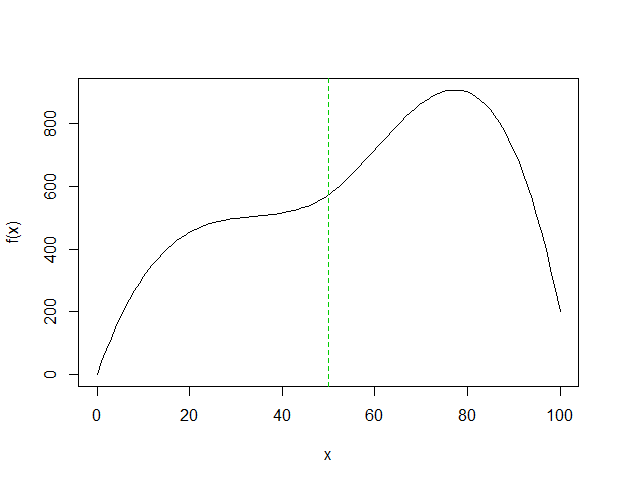}}
\subfigure[$t_1 = 25$ and $t_2 = 75$.]{
\includegraphics[width=5cm]{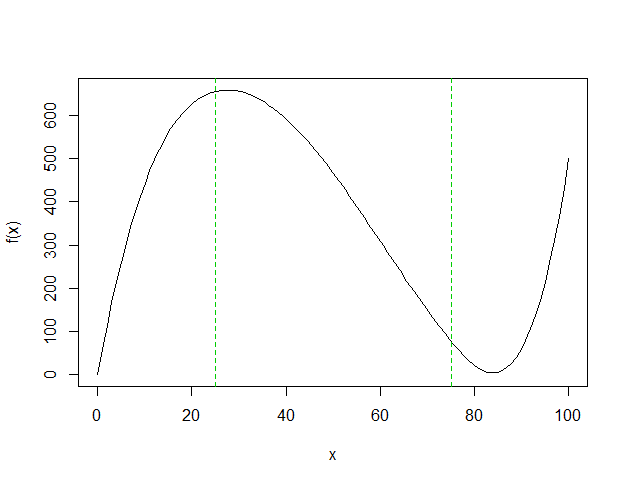}}
\subfigure[$t_1=25$, $t_2=50$ and $t_3=75$.]{
\includegraphics[width=5cm]{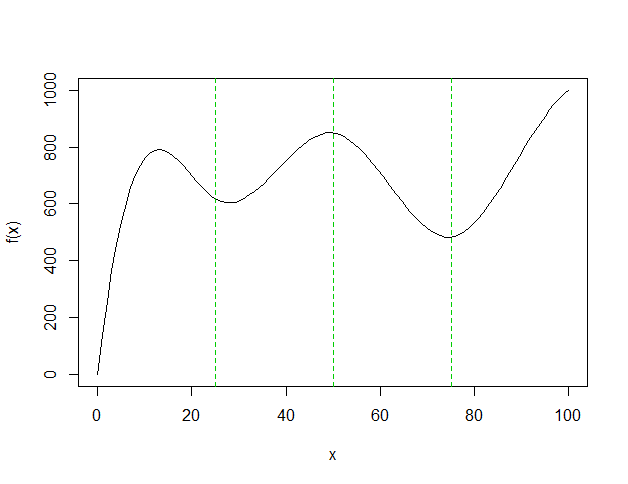}}
\caption{The solid black lines are the underlying cubic B-spline functions considered in the simulation studies and the green dashed vertical lines indicate the knots locations.}\label{fig:spline}
\end{figure}

We apply the proposed method in synthetic data to evaluate its performance in several scenarios.
Figure~\ref{fig:spline} shows the cubic B-splines underlying functions considered: in the left panel with a knot at $t_1=50$, in the central panel with two knots located at $t_1=25$ and $t_2=75$, and the right panel with three knots at $t_1=25$, $t_2=50$, and $t_3=75$.

We take samples from each of the tree functions with additional 
zero mean normal noise with variance according to three signal to noise ratio values (SNR), SNR = 3, 6, and 9 and two sample sizes, $n=100$, and $1000$. 
We apply the proposed method to estimate the number and location of the knots considering $\delta =15$ for minimum knots distance.
This process was repeated $1000$ times for each scenario to obtain the 
mean, median, standard deviation (SD) of the estimated knots, also the proportion of correctly estimated number of knots and the $95\%$ confidence interval (CI). 
Tables \ref{tab:1knot}, \ref{tab:2knots}, and \ref{tab:3knots} show the proportion of correctly estimated number of knots for the three underlying functions in Figure~\ref{fig:spline}. 

\begin{figure}[H]
\centering
\subfigure[1 knot, SNR=3]{
\includegraphics[width=5cm]{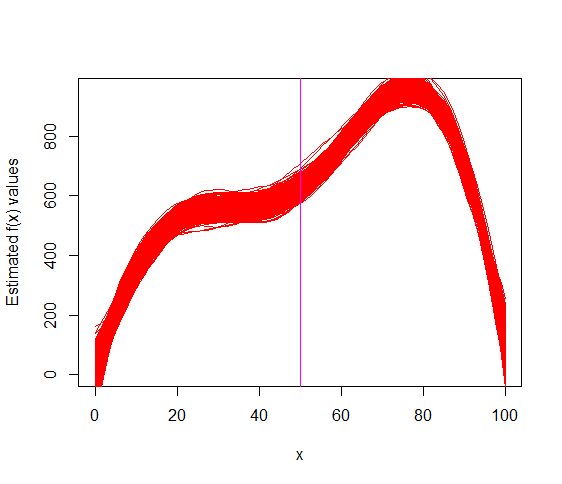}}
\subfigure[1 knot, SNR=6]{
\includegraphics[width=5cm]{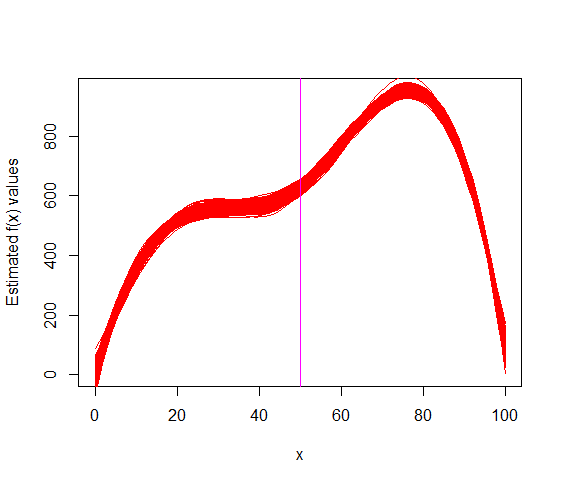}}
\subfigure[1 knot, SNR=9]{
\includegraphics[width=5cm]{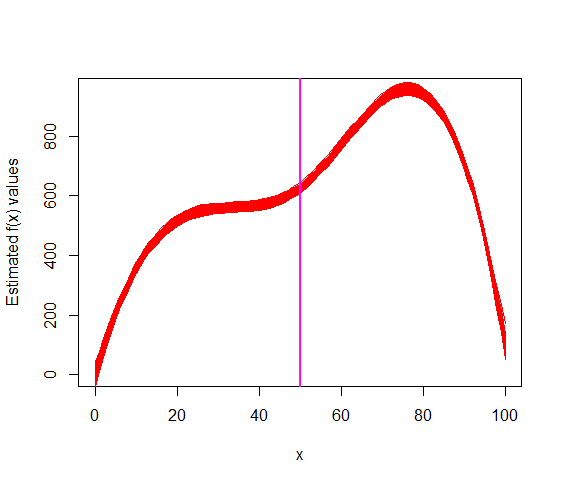}}
\subfigure[2 knots, SNR=3]{
\includegraphics[width=5cm]{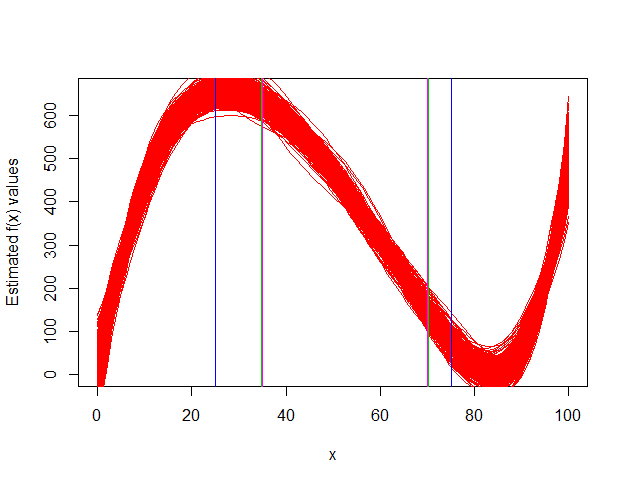}}
\subfigure[2 knots, SNR=6]{
\includegraphics[width=5cm]{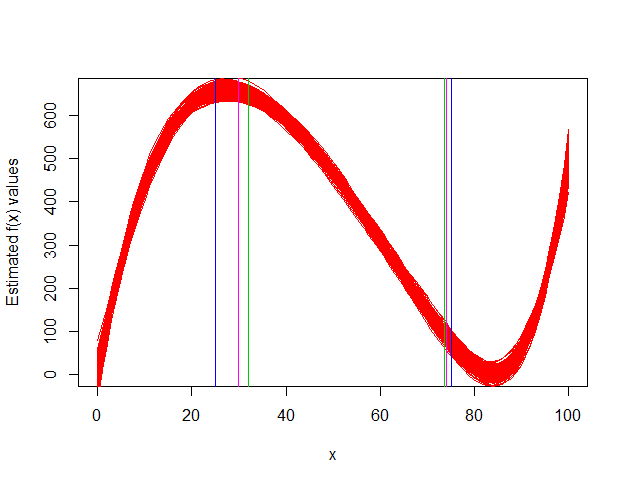}}
\subfigure[2 knots, SNR=9]{
\includegraphics[width=5cm]{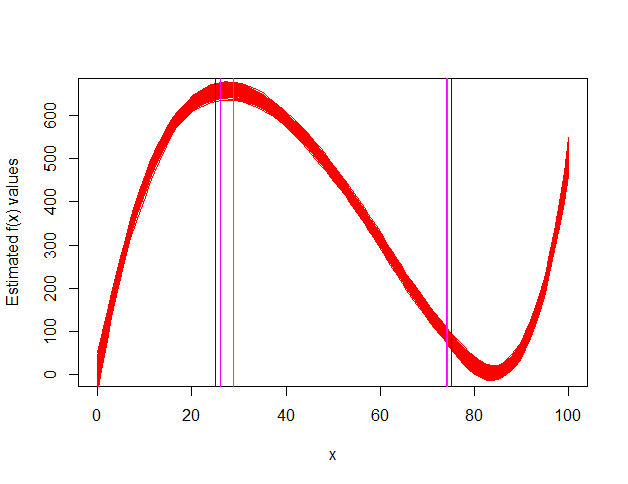}}
\subfigure[3 knots, SNR=3]{
\includegraphics[width=5cm]{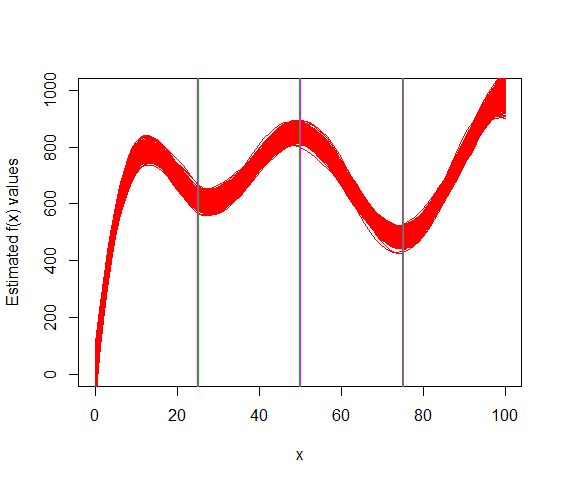}}
\subfigure[3 knots, SNR=6]{
\includegraphics[width=5cm]{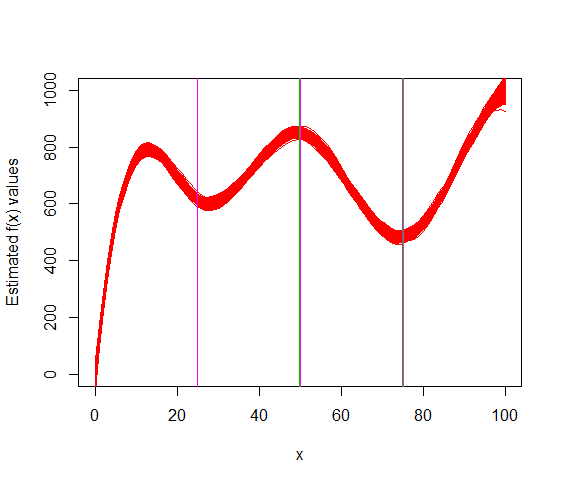}}
\subfigure[3 knots, SNR=9]{
\includegraphics[width=5cm]{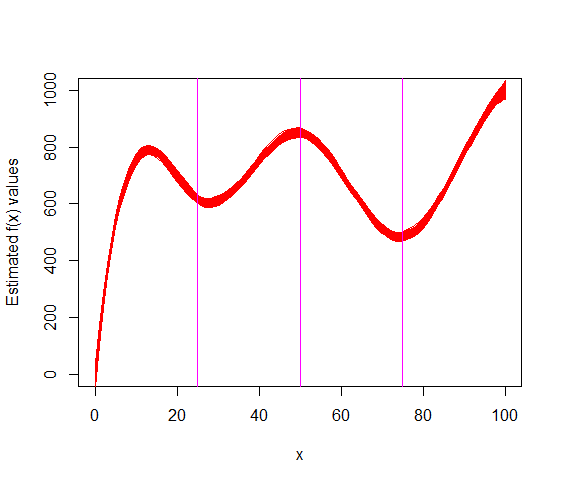}}
\caption{Estimated curves for $n=100$. Blue vertical lines indicate the true knots locations. Green and pink vertical lines are the mean and median of the estimated knots locations, respectively. Note that in most of the scenarios, the vertical lines are overlapping.}\label{fig:fit}
\end{figure}

\begin{figure}[H]
\centering
\subfigure[1 knot, SNR=3]{
\includegraphics[width=5cm]{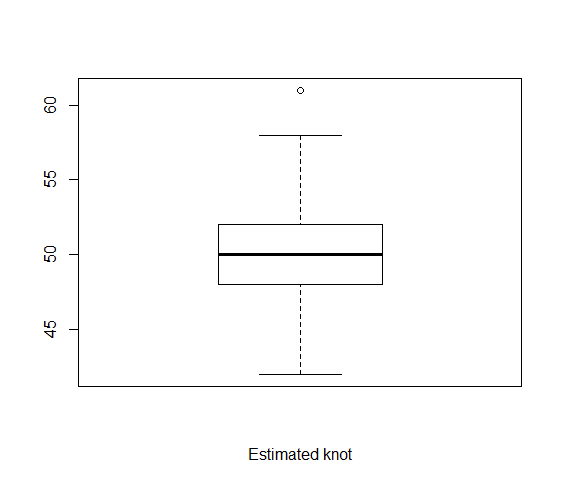}}
\subfigure[1 knot, SNR=6]{
\includegraphics[width=5cm]{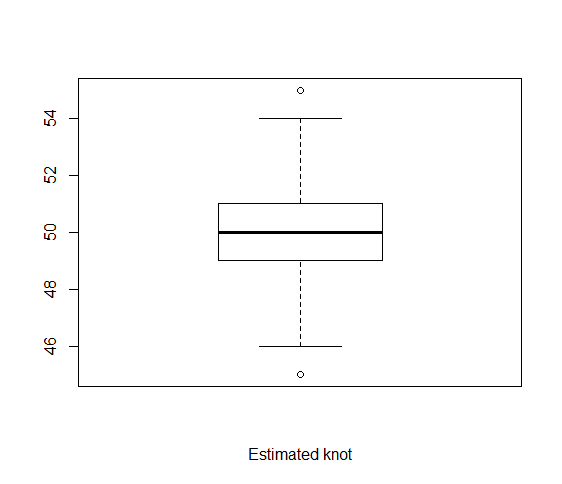}}
\subfigure[1 knot, SNR=9]{
\includegraphics[width=5cm]{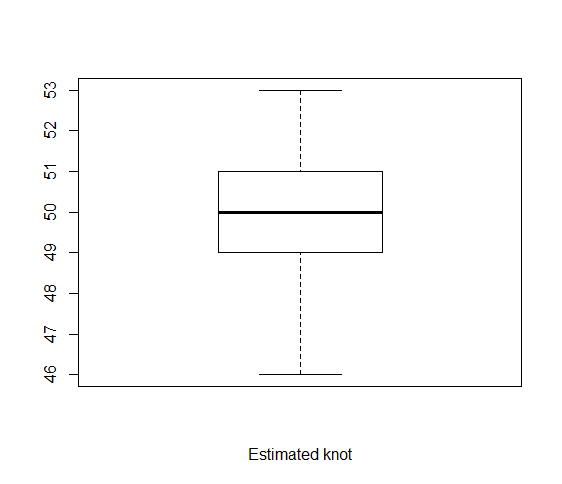}}
\subfigure[2 knots, SNR=3]{
\includegraphics[width=5cm]{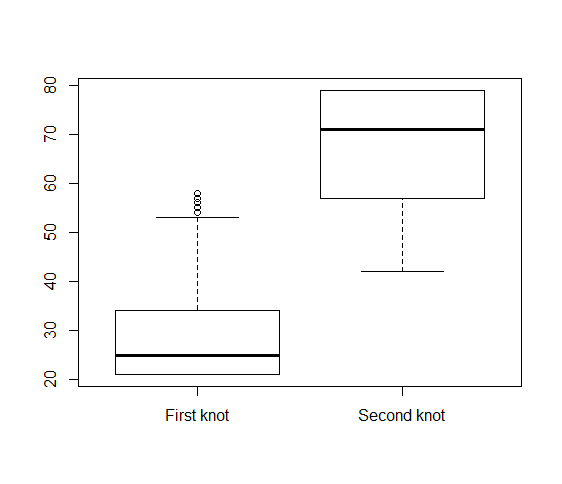}}
\subfigure[2 knots, SNR=6]{
\includegraphics[width=5cm]{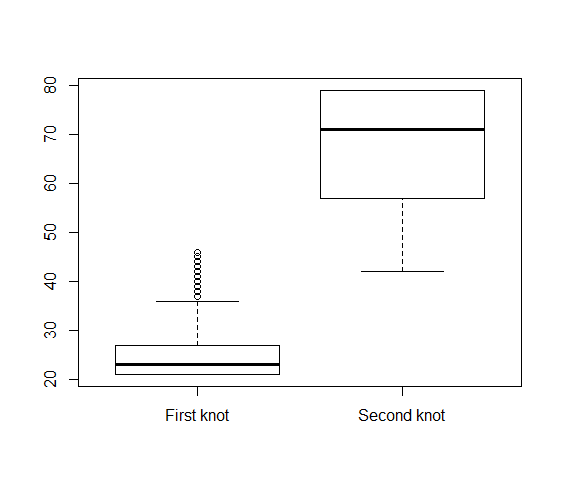}}
\subfigure[2 knots, SNR=9]{
\includegraphics[width=5cm]{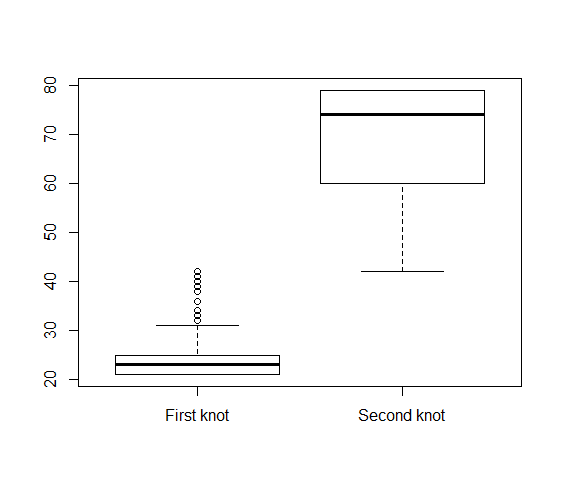}}
\subfigure[3 knots, SNR=3]{
\includegraphics[width=5cm]{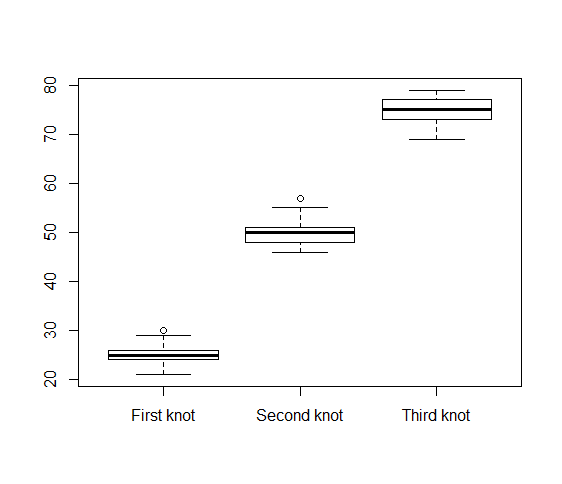}}
\subfigure[3 knots, SNR=6]{
\includegraphics[width=5cm]{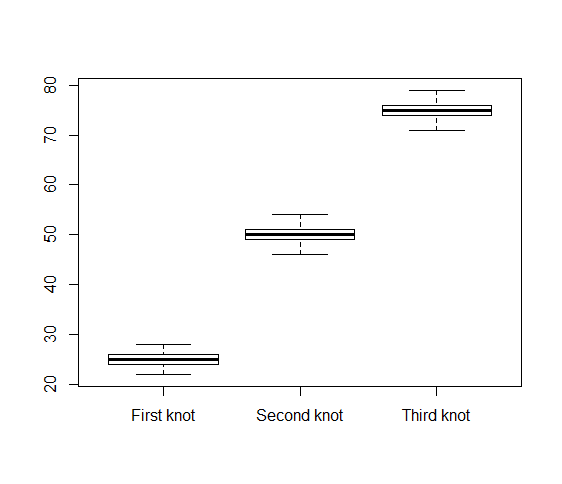}}
\subfigure[3 knots, SNR=9]{
\includegraphics[width=5cm]{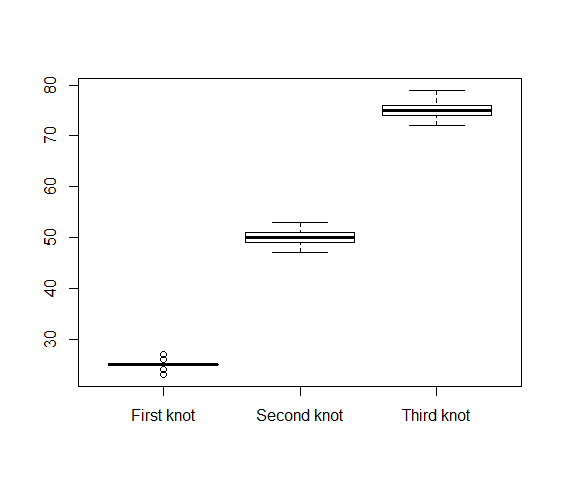}}
\caption{Boxplots of the estimated knots for $n=100$.}\label{fig:bp}
\end{figure}

\begin{figure}[H]
\centering
\subfigure[SNR=3, 1st knot.]{
\includegraphics[width=5cm]{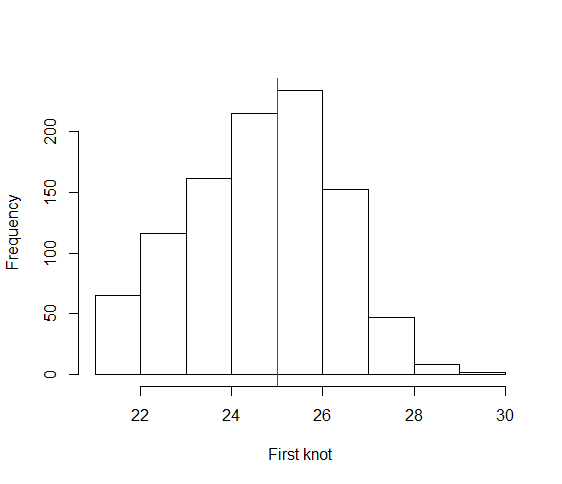}}
\subfigure[SNR=3, 2nd knot.]{
\includegraphics[width=5cm]{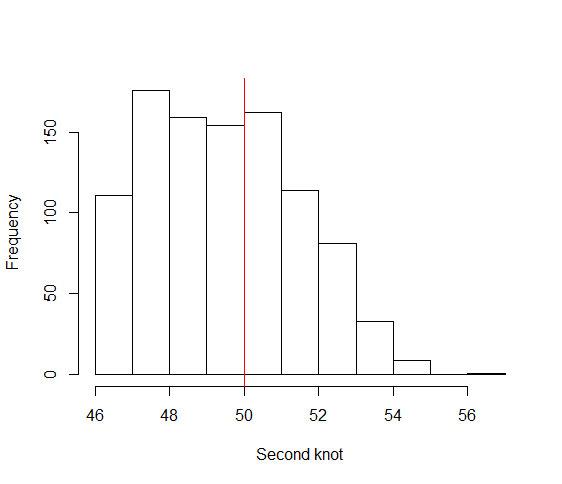}}
\subfigure[SNR=3, 3rd knot.]{
\includegraphics[width=5cm]{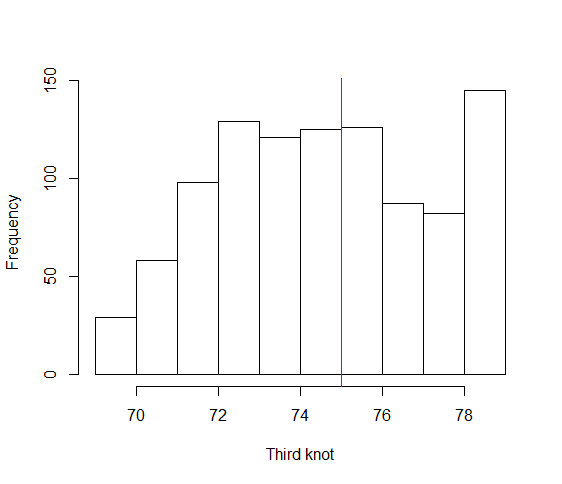}}
\subfigure[SNR=6, 1st knot.]{
\includegraphics[width=5cm]{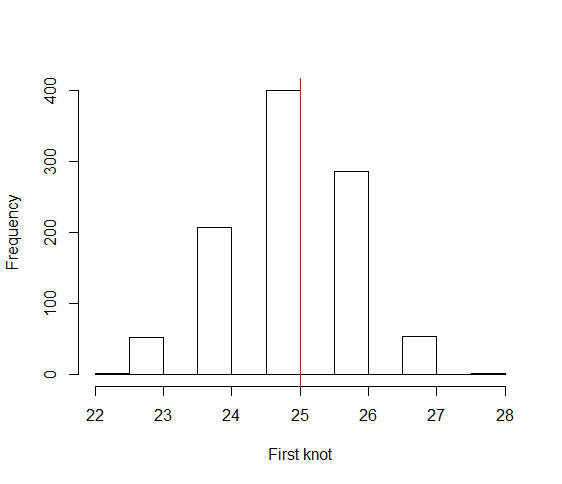}}
\subfigure[SNR=6, 2nd knot.]{
\includegraphics[width=5cm]{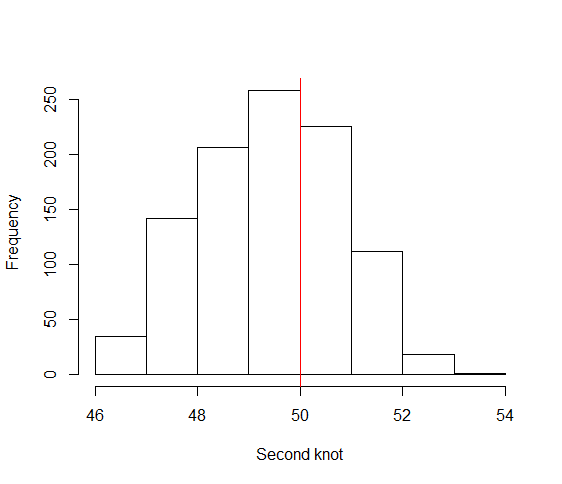}}
\subfigure[SNR=6, 3rd knot.]{
\includegraphics[width=5cm]{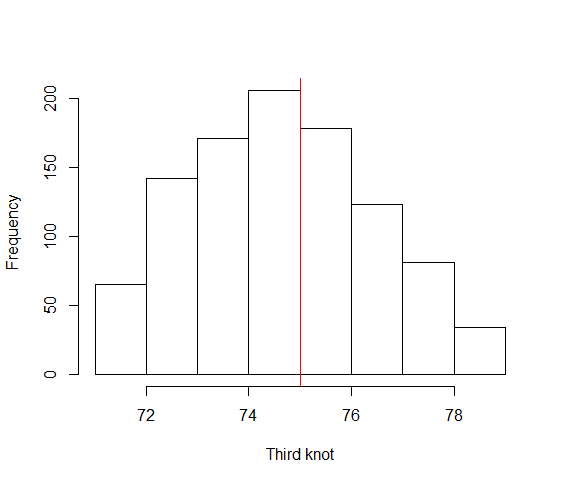}}
\subfigure[SNR=9, 1st knot.]{
\includegraphics[width=5cm]{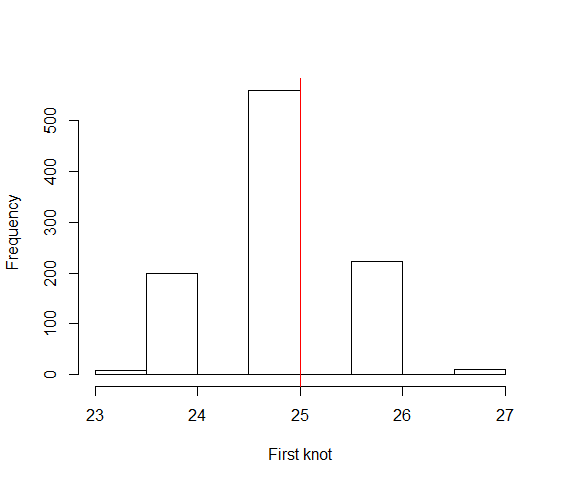}}
\subfigure[SNR=9, 2nd knot.]{
\includegraphics[width=5cm]{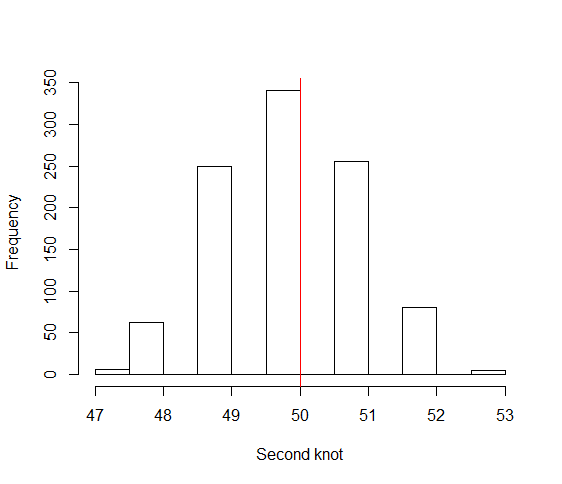}}
\subfigure[SNR=9, 3rd knot.]{
\includegraphics[width=5cm]{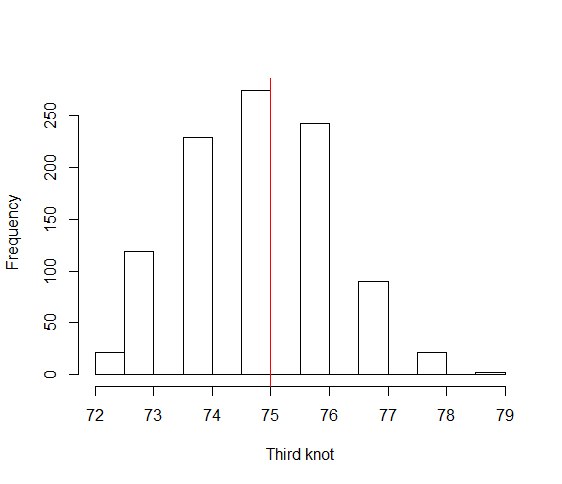}}
\caption{Histograms of the estimated knots for $n=100$ and $K=3$ knots.}\label{fig:hist}
\end{figure}

In fact, our method accurately estimated the number and location of the knots in the majority of the scenarios considered.
Even for low SNR value (SNR=3), 
the estimates were very close to the true knots location with low standard deviation values. As expected, estimates improved as SNR and/or $n$ increased. The curve estimates with the true values of the knots, mean and median values for the estimated knots are shown in Figure \ref{fig:fit}, boxplots and histograms of the estimated knots are presented in Figures \ref{fig:bp} and \ref{fig:hist}, respectively, for $n=100$. The behavior of the estimates for $n=1000$ was similar.

\begin{table}[H]
\scalefont{0.9}
\centering
\label{my-label}
\begin{tabular}{|c|c|c|c|c|c|c|c|}
\hline 
\textbf{n} & \textbf{SNR} & $\boldsymbol{\%}$ of $\boldsymbol{\hat{K}=1}$& \textbf{Knot} & \textbf{Mean} & \textbf{Median} & \textbf{SD} & \textbf{CI(95$\%$)}   \\ \hline \hline
100 & 3 & 77$\%$ &1 & 50.18 & 50.00 & 3.19 & (44.00 ; 56.00) \\ \hline
    & 6 & 81$\%$ &1 & 49.99 & 50.00 & 1.61 & (47.00 ; 53.00) \\ \hline
    & 9 & 99$\%$& 1 & 49.92 & 50.00 & 1.13  & (48.00 ; 52.00) \\ \hline \hline
1000& 3 & 100$\%$& 1 & 49.93 & 50.00 & 1.23 & (48.00 ; 52.00) \\ \hline
    & 6 & 100$\%$ &1 & 50.05 & 50.00 & 0.63 & (49.00 ; 51.00) \\ \hline
    & 9 & 100$\%$ &1 & 49.99 & 50.00 & 0.48 & (49.00 ; 51.00) \\ \hline
\end{tabular}
\caption{Simulation results of the proposed method for data generated from a cubic B-splines with one knot at $t_1=50$.}\label{tab:1knot}
\end{table}

\begin{table}[H]
\scalefont{0.9}
\centering
\label{my-label}
\begin{tabular}{|c|c|c|c|c|c|c|c|}
\hline 
\textbf{n} & \textbf{SNR} & $\boldsymbol{\%}$ of $\boldsymbol{\hat{K}=2}$& \textbf{Knot} & \textbf{Mean} & \textbf{Median} & \textbf{SD} & \textbf{CI(95$\%$)}   \\ \hline \hline
100 & 3 & 73$\%$ &1 & 28.90 & 25.00 & 9.63 & (21.00 ; 56.00) \\
    &   & &2 & 67.11 & 71.00 & 12.32 & (42.00 ; 79.00) \\ \hline
    & 6 & 84$\%$ &1 & 24.62 & 23.00 & 4.81 & (21.00 ; 37.65) \\
    &   & &2 & 67.45 & 71.00 & 12.01 & (42.35 ; 79.00) \\ \hline
    & 9 & 91$\%$ &1 & 23.63 & 23.00 & 3.16  & (21.00 ; 31.00) \\
    &   & &2 & 68.68 & 74.00 & 11.55 & (44.00 ; 79.00) \\ \hline \hline
1000& 3 & 73$\%$ &1 & 27.95 & 26.00 & 7.55 & (21.00 ; 46.43) \\
    &   &  &2 & 74.53 & 75.00 & 2.78 & (68.00 ; 79.00) \\ \hline
    & 6 & 86$\%$ &1 & 25.73 & 25.00 & 3.91 & (21.00 ; 35.00) \\
    &   & &2 & 74.80 & 75.00 & 1.27 & (72.00 ; 77.00) \\ \hline
    & 9 & 95$\%$ &1 & 25.17 & 25.00 & 2.55 & (21.00 ; 30.00) \\
    &   & &2 & 74.93 & 75.00 & 0.92 & (73.00 ; 77.00) \\ \hline     
\end{tabular}
\caption{Simulation results of the proposed method for data generated from a cubic B-splines with two knots at $t_1=25$ and $t_2=75$.}\label{tab:2knots}
\end{table}

\begin{table}[H]
\scalefont{0.9}
\centering
\label{my-label}
\begin{tabular}{|c|c|c|c|c|c|c|c|}
\hline 
\textbf{n} & \textbf{SNR} & $\boldsymbol{\%}$ of $\boldsymbol{\hat{K}=3}$& \textbf{Knot} & \textbf{Mean} & \textbf{Median} & \textbf{SD} & \textbf{CI(95$\%$)}   \\ \hline \hline
100 & 3 & 100$\%$ & 1 & 25.11 & 25.00 & 1.67 & (22.00 ; 28.00) \\  
    &   &  &2 & 49.94 & 50.00 & 2.06 & (46.00 ; 54.00) \\ 
    &   &  &3 & 75.07 & 75.00 & 2.60 & (70.00 ; 79.00) \\ \hline
  & 6 & 100$\%$&1 & 25.08 & 25.00 & 0.96 & (23.00 ; 27.00) \\
  &   & & 2 & 49.91 & 50.00 & 1.39 & (47.00 ; 52.00) \\
  &   & &3 & 75.14 & 75.00 & 1.83 & (72.00 ; 79.00) \\ \hline
  & 9 & 100$\%$& 1 & 25.03 & 25.00 & 0.71 & (24.00 ; 26.00) \\
  &   & &2 & 50.04 & 50.00 & 1.09 & (48.00 ; 52.00) \\
  &   & &3 & 74.96 & 75.00 & 1.31 & (73.00 ; 77.00) \\ \hline \hline
1000 & 3 &100$\%$ & 1 & 25.07 & 25.00 & 0.67 & (24.00 ; 26.00) \\
     &   & & 2 & 49.92 & 50.00 & 1.08 & (48.00 ; 52.00) \\
     &   & & 3 & 75.07 & 75.00 & 1.26 & (73.00 ; 77.00) \\ \hline
     & 6 & 100$\%$ & 1 & 25.02 & 25.00 & 0.27 & (24.48 ; 26.00) \\
     &   & &2 & 49.97 & 50.00 & 0.49 & (49.00 ; 51.00) \\
     &   & & 3 & 75.08 & 75.00 & 0.66 & (74.00 ; 76.00) \\ \hline
     & 9 & 100$\%$&1 & 25.00 & 25.00 & 0.10 & (24.80 ; 25.30) \\
     &   & &2 & 50.01 & 50.00 & 0.23 & (50.00 ; 51.00) \\
     &   & &3 & 74.99 & 75.00 & 0.34 & (74.00 ; 76.00) \\ \hline
\end{tabular}
\caption{Simulation results of the proposed method for data generated from a cubic B-splines with three knots at $t_1=25$, $t_2=50$ and $t_3=75$.}\label{tab:3knots}
\end{table}

\subsection{Simulation study 2} \label{sec:sim2}
We now compare the performance of our method with two methods that also automatically select the knots, namely P-spline, or Penalized Spline (Ruppert, 2002; and Ruppert et al., 2003) and the recently proposed A-spline, Adaptive Spline (Goepp et al., 2018).





\begin{figure}[H]
\centering
\subfigure[SpaHet function.]{
\includegraphics[scale=0.4]{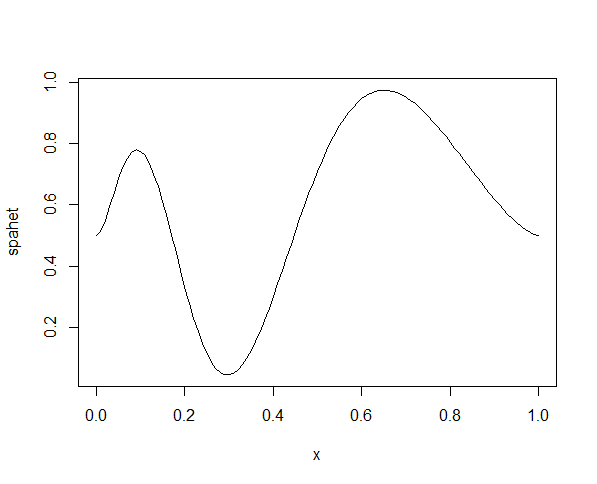}}
\subfigure[Sine function.]{
\includegraphics[scale=0.4]{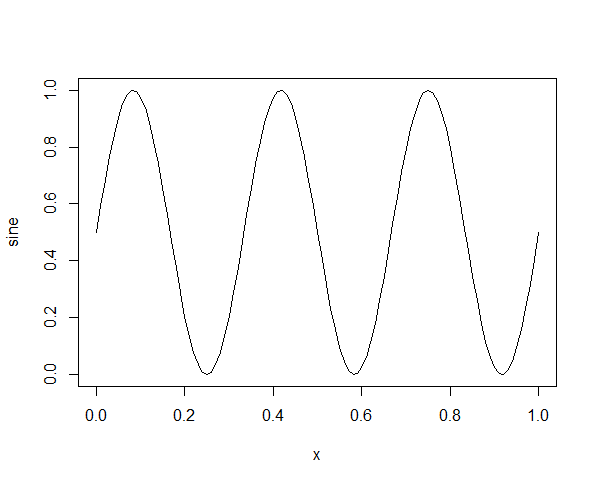}}
\subfigure[Logit function.]{
\includegraphics[scale=0.4]{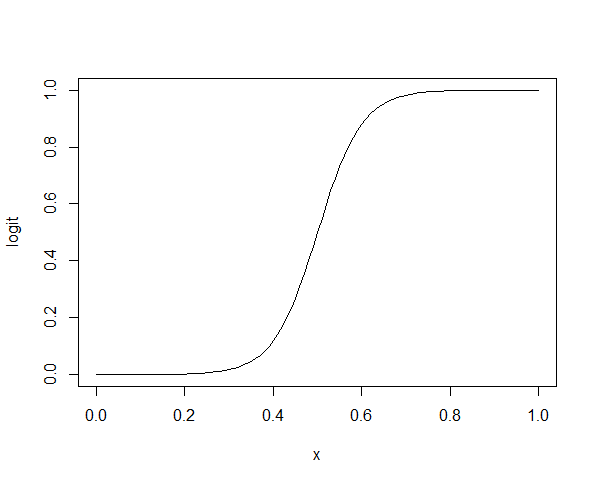}}
\caption{Graph of the underlying functions 
used for data generation.} \label{fig:curves21}
\end{figure}

The data was generated from three commonly used  functions in spline regression simulation studies: SpaHet (for spatially heterogeneous), Sine and Logit, see Wand (2000), Ruppert (2002) and Goepp et al. (2018).
Their definition are in Table~\ref{tab:functions} and their graph can be seen in Figure~\ref{fig:curves21}.

\begin{table}[]
\centering
\begin{tabular}{@{}ll@{}}
\toprule
 \multicolumn{1}{c}{Function name} & \multicolumn{1}{c}{Function definition} \\ \midrule
 SpaHet & $f(x) = \sqrt{x(1-x)}\sin\left(\frac{2\pi(1+2^{-0.6})}{x+2^{-0.6}}\right)$ \\
 Sine & $f(x) = 0.5\sin(6\pi x) + 0.5$ \\
 Logit & $f(x) = \frac{1}{1+\exp\{-20(x-0.5)\}}$ \\ \bottomrule
\end{tabular}
\caption{Functions used for data generation.}
\label{tab:functions}
\end{table}

\begin{figure}[H]
\centering
\subfigure[SpaHet function.]{
\includegraphics[scale=0.4]{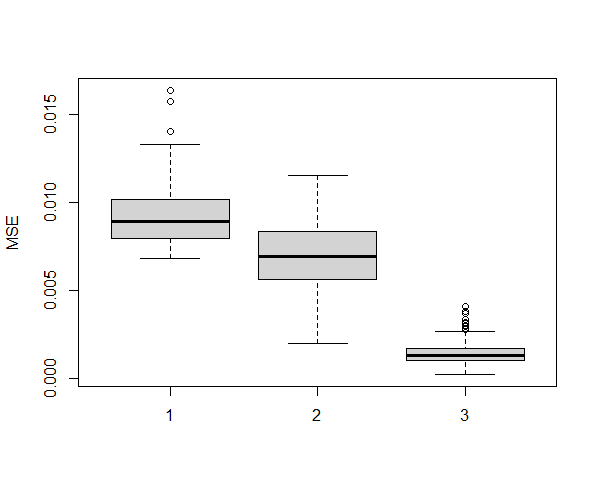}}
\subfigure[Sine function.]{
\includegraphics[scale=0.4]{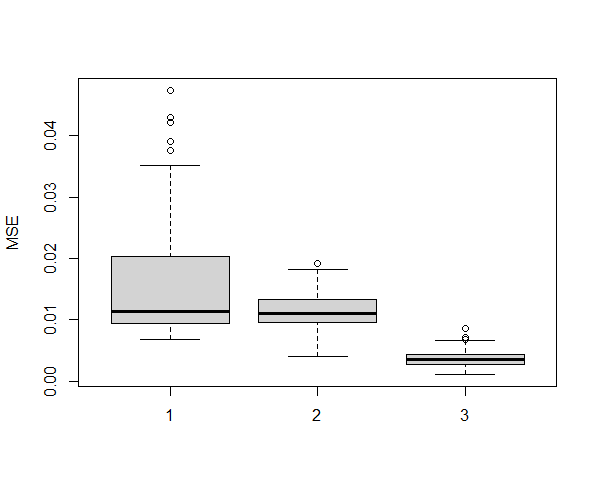}}
\subfigure[Logit function.]{
\includegraphics[scale=0.4]{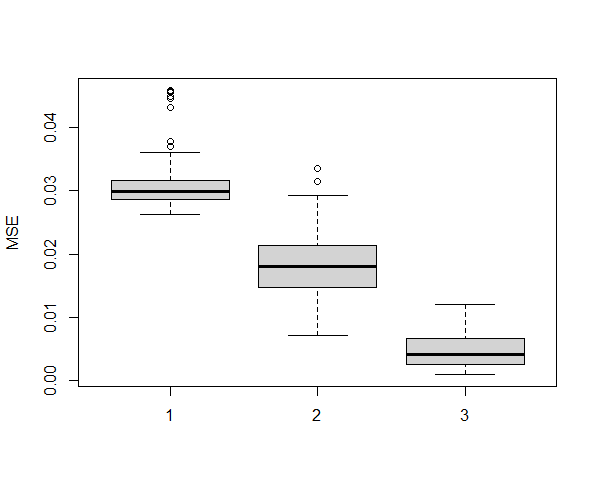}}
\caption{Boxplots of MSEs for the penalized spline regression methods in simulation study 2 for $n=50$ and SNR = 3. Associated methods: 1-P-spline, 2-A-spline and 3-Proposed method.} \label{fig:bp21}
\end{figure}

For each of the underlying functions considered, we generate equally spaced points for two scenarios with different sample sizes: $n=10$ and $n=100$.
Moreover, we added to each observation a zero mean Gaussian random noise with variance related to the signal to noise ratio values SNR=3, SNR=9.
Then, for each combination of the underlying function, sample size and SNR we generated $M=200$ replicates.
We computed the mean squared error (MSE)
$$MSE = \frac{1}{n} \sum_{i=1}^{n}[{\hat f(x_i)} - f(x_i)]^2,$$
for each replicate and as a performance metric, we considered the averaged mean squared error (AMSE), $$AMSE = \frac{1}{M} \sum_{j=1}^{M}MSE_j.$$
Table \ref{tab:sim21} presents the AMSE and the standard deviation (SD) of the MSEs for the scenarios considered and Figure \ref{fig:bp21} shows boxplots of the MSE for $n=50$ and SNR=3.

\begin{table}[H]
\centering
\label{my-label}
\begin{tabular}{|c|c|c|c|c|}
\hline
\textbf{Function} & \textbf{n} & \textbf{Method} & \textbf{SNR=3} & \textbf{SNR=9}  \\ \hline 
   &  &        & \textbf{AMSE (SD)} & \textbf{AMSE (SD)}  \\ \hline \hline
SpaHet& 50	&	P-spline 	&	92.86 (174.68) & 75.01 (62.33) \\
      &      &   A-spline     &  69.80 (180.41) & 4.43 (32.11) \\
      &      &   Proposed Method & \textbf{14.93 (71.08)} & \textbf{2.33 (8.89)} \\ \hline
      & 100	&	P-spline	&	10.44 (42.41) & 1.49 (5.48) \\
      &      &   A-spline     &  15.52 (89.25) & \textbf{1.41 (6.33)} \\
      &      &   Proposed Method & \textbf{7.99 (36.06)} & 2.03 (4.20)	\\ \hline \hline

 Sine & 50	&	P-spline 	&	153.72 (847.06) & 87.49 (402.29)\\
      &      &   A-spline     &   113.24 (294.26) & \textbf{9.21 (44.63)}\\
      &      &   Proposed Method &  \textbf{36.97 (130.47)} & 10.64 (17.35)\\ \hline
      & 100	&	P-spline 	&	\textbf{20.75 (80.62)} & \textbf{2.89 (9.82)}\\
      &      &   A-spline    &   34.00 (139.21) & 3.61 (1.24) \\
      &      &   Proposed Method & 	21.70 (67.36) & 7.65 (8.59) \\ \hline \hline

Logit & 50	&	P-spline 	& 306.27 (333.29) & 280.58 (59.87)	\\
      &      &   A-spline     &  181.09 (488.69)& 15.00 (82.24)\\
      &      &   Proposed Method & \textbf{47.07 (273.35)} & \textbf{6.18 (59.62)} \\ \hline
      & 100	&	P-spline 	& \textbf{18.63	(99.96)} & \textbf{2.73 (12.36)} \\
      &      &   A-spline    & 38.51 (240.85)  & 4.20 (21.38) \\
      &      &   Proposed Method & 21.28 (104.36)& 4.08 (9.54) 	 \\ \hline

\end{tabular}
\caption{AMSEs ($\times 10^{-4}$) and their standard deviations ($\times 10^{-5}$) of our proposed method, P-spline, and A-spline 
considering the underlying functions SpaHet, Sine, and Logit 
}\label{tab:sim21}
\end{table}

\begin{figure}[H]
\centering
\subfigure[SpaHet function estimates.]{
\includegraphics[scale=0.4]{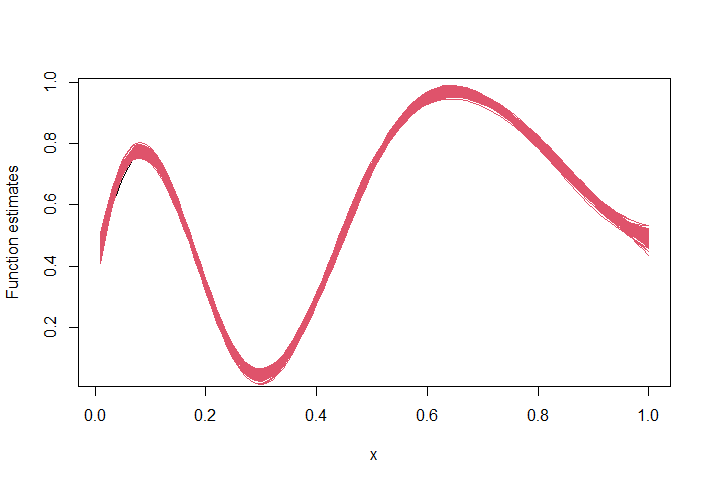}}
\subfigure[Sine function estimates.]{
\includegraphics[scale=0.4]{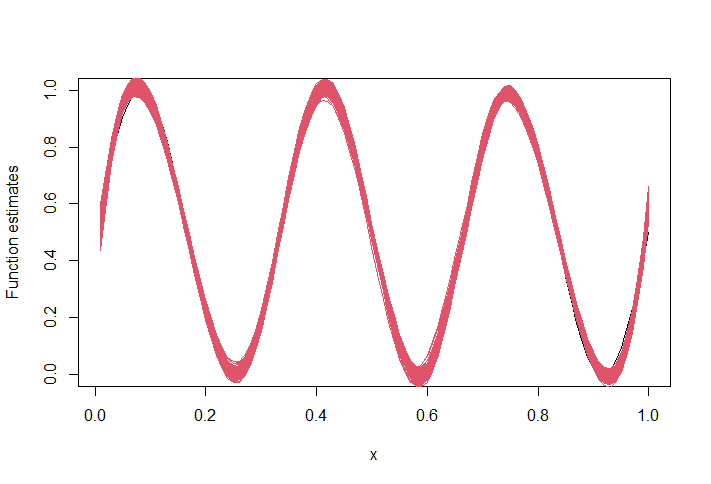}}
\subfigure[Logit function estimates.]{
\includegraphics[scale=0.4]{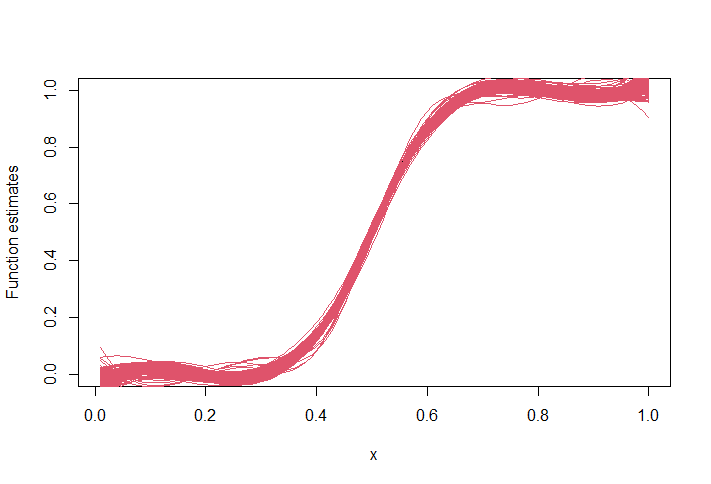}}
\caption{Estimated curves obtained by the proposed method in simulation study 2 for $n=100$ and SNR=9.} \label{fig:curves22}
\end{figure}

In general, the proposed method 
presented a competing performance in all scenarios considered.
Note that in scenarios with more noise levels, SNR=3, our method outperformed the A-spline and P-spline in four out of six cases.
Furthermore, 
our method had better performance in scenarios with sample size $n = 50$.
Finally, even when the method was not the best, especially for SNR=9, its performance was quite close to the competing methods. 
Thus, our simulation study indicates that the proposed method can be considered by practitioners in application in real data, with a special advantage in data with high noise level. 
Figure \ref{fig:curves22} shows the curve estimates by the method for $n=100$ and SNR=9. 
Note that the estimates recovered well the main features of the underlying functions, such as peaks and local maximum and minimum points.

\section{Analysis of Covid-19 time series}
The counts of daily new cases of the novel coronavirus disease (Covid-19) varies considerably 
throughout the affected countries and regions, mainly due to testing criteria used, delays in 
notification, and population and authorities response to the outbreak, among others.
Even though, the analysis of the disease's daily new infected cases is important to understand at 
what stage of the epidemic each country is in.
Therefore, we can use the method presented above to estimate the dates at which the trend of new 
cases counts had changed.
This may be helpful not only to define public policy such as when safely relax social distance 
measures but also to verify the effectiveness of mitigations measures taken.
Here we perform estimation and prediction for Covid-19 data using our proposed method, the data is 
publicly available at \url{https://opendata.ecdc.europa.eu/covid19/}.

In general, during an epidemic the daily number of new infected cases shows several trends that are 
not related to the disease spreading itself, but to other causes, such as delay in test processing time 
or lack of testing. For this reason, we considered the 7-days moving average smoother on the time 
series before applying our method. Further, to avoid knots estimation in the beginning of the time 
series, we remove the 10\% initial dates as possible knots locations.
 
\begin{figure}[H]
\centering
\subfigure[Brazil.\label{lognormal}]{
\includegraphics[width=5cm]{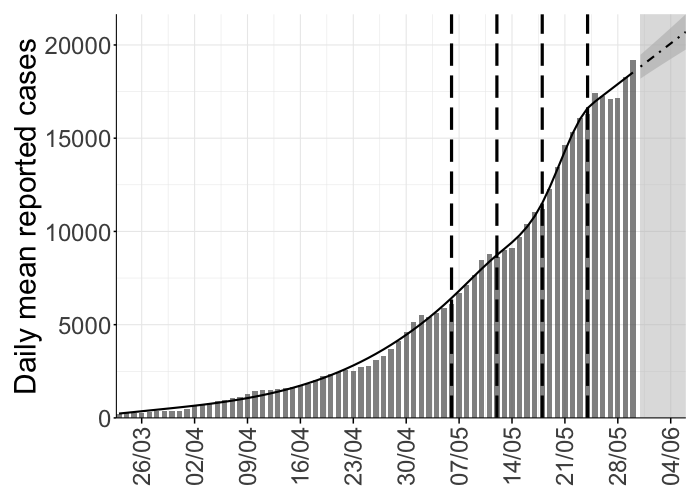}}
\subfigure[USA.\label{blocls}]{
\includegraphics[width=5cm]{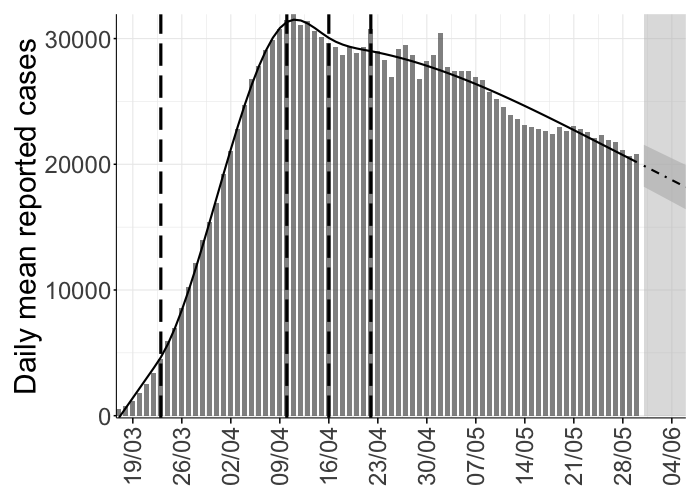}}
\subfigure[Spain.\label{blocls}]{
\includegraphics[width=5cm]{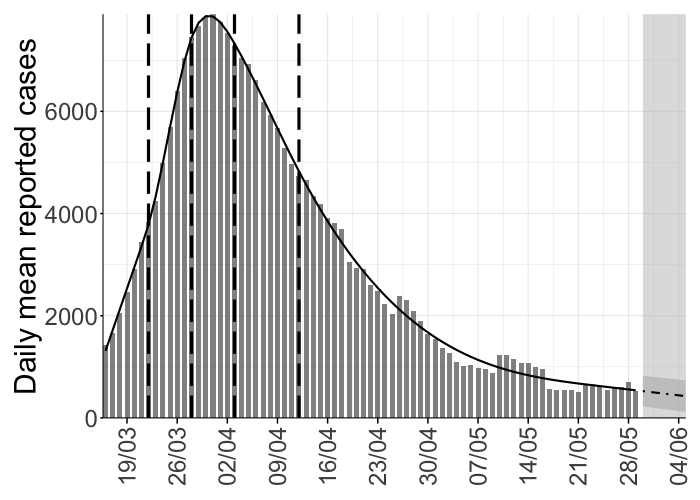}}
\subfigure[South Korea.\label{blocls}]{
\includegraphics[width=5cm]{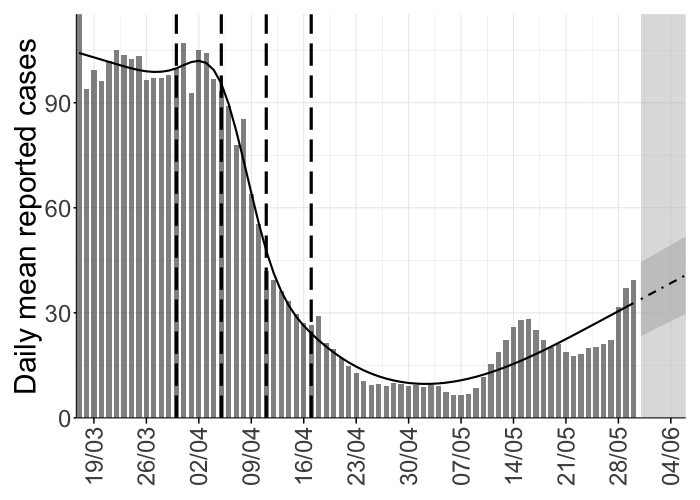}}
\subfigure[Iran.\label{blocls}]{
\includegraphics[width=5cm]{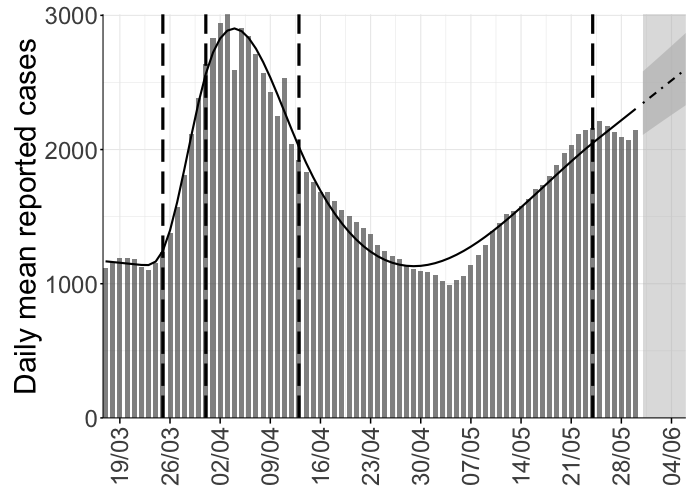}}
\subfigure[Switzerland.\label{blocls}]{
\includegraphics[width=5cm]{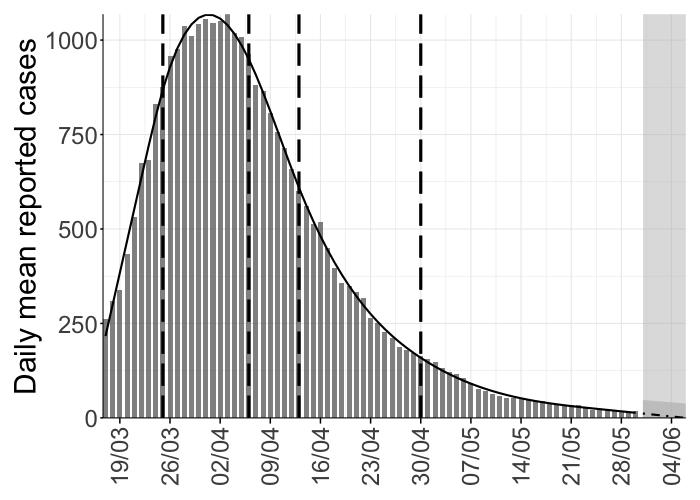}}
\subfigure[Germany.\label{blocls}]{
\includegraphics[width=5cm]{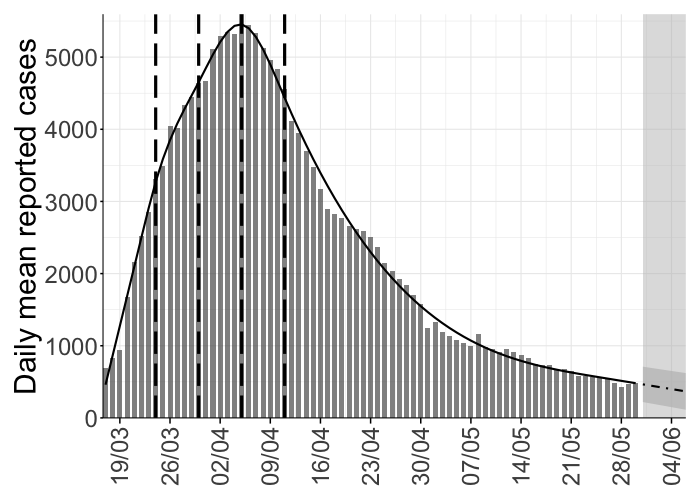}}
\subfigure[Italy.\label{blocls}]{
\includegraphics[width=5cm]{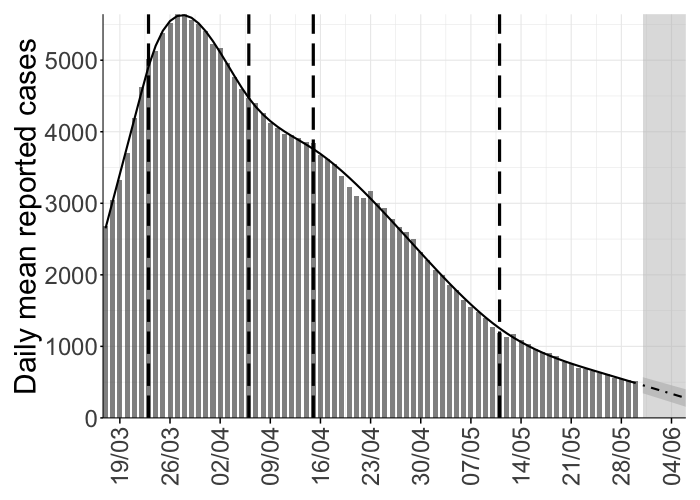}}
\caption{Natural cubic splines curve fitting of mean daily new registered cases of Covid-19 in linear scale for eight countries considering data until May 30th 2020. In each graph, the gray bars indicate mean daily recorded new cases, the solid black line is the fitted curve, dotted black line is the prediction for the following 7 days, the shaded area indicates the interval of estimates, vertical dashed lines represent the knots locations.
}\label{fig:covidcubico}
\end{figure}

\begin{figure}[H]
\centering
\subfigure[Brazil.\label{}]{
\includegraphics[width=5cm]{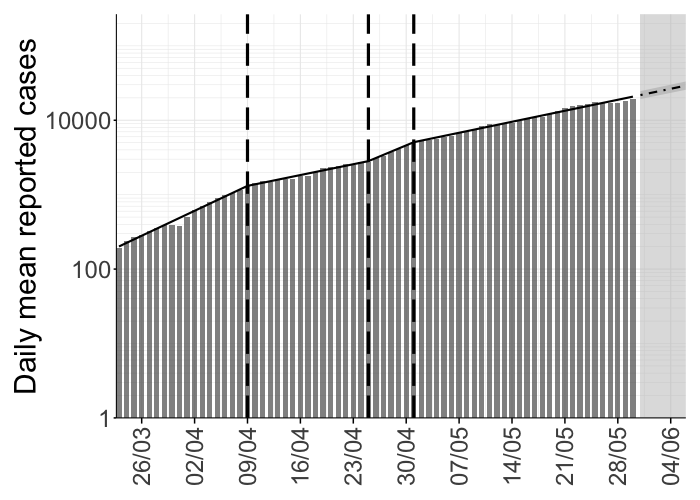}}
\subfigure[USA.\label{blocls}]{
\includegraphics[width=5cm]{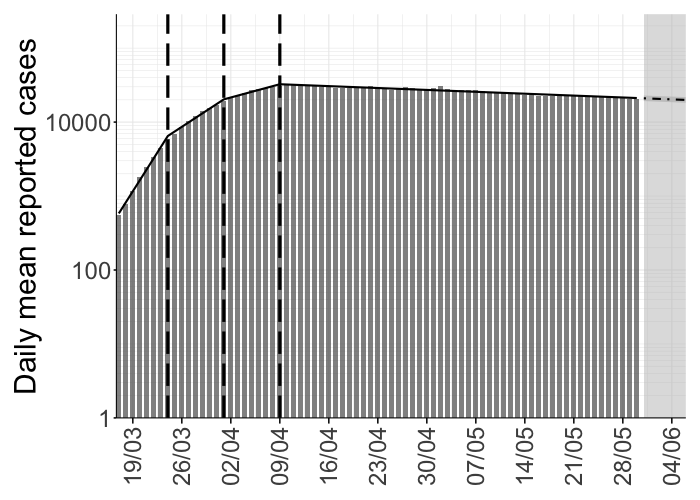}}
\subfigure[Spain.\label{}]{
\includegraphics[width=5cm]{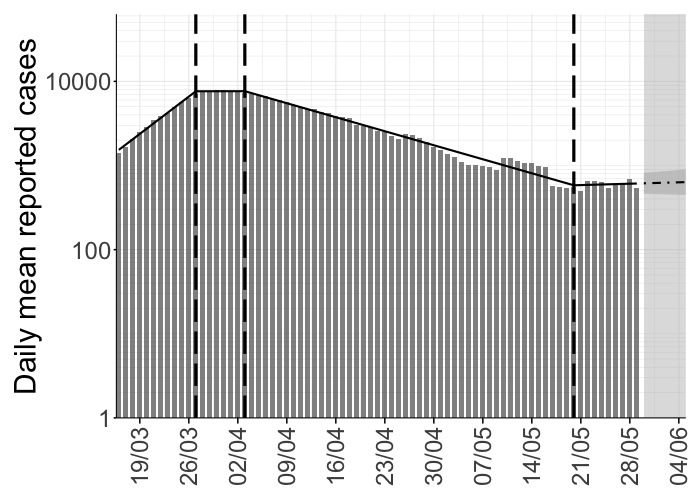}}
\subfigure[South Korea.\label{blocls}]{
\includegraphics[width=5cm]{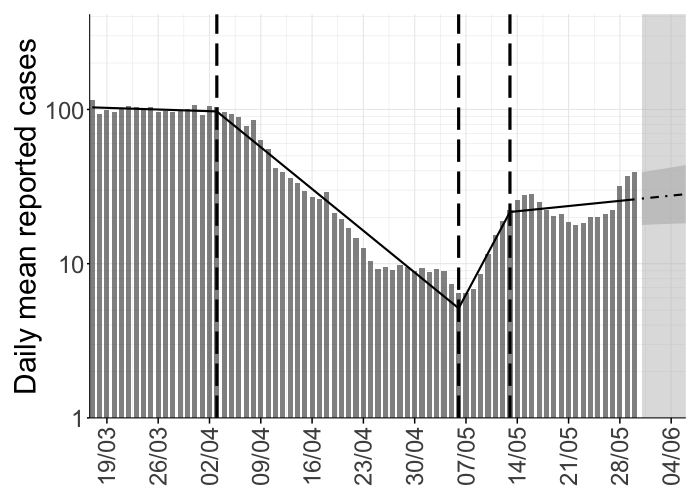}}
\subfigure[Iran.\label{}]{
\includegraphics[width=5cm]{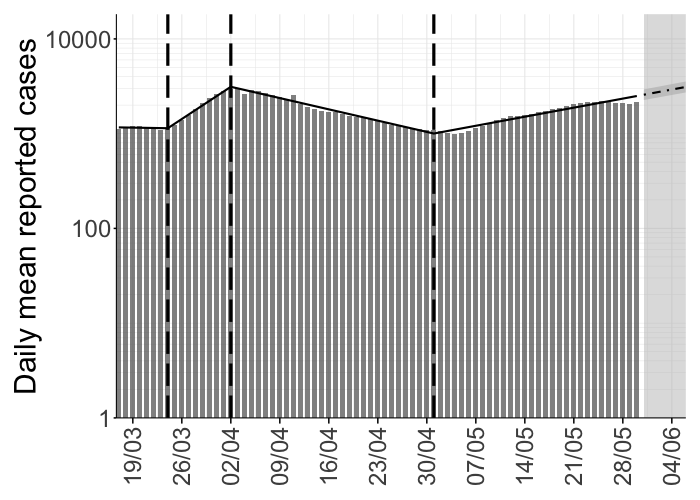}}
\subfigure[Switzerland.\label{}]{
\includegraphics[width=5cm]{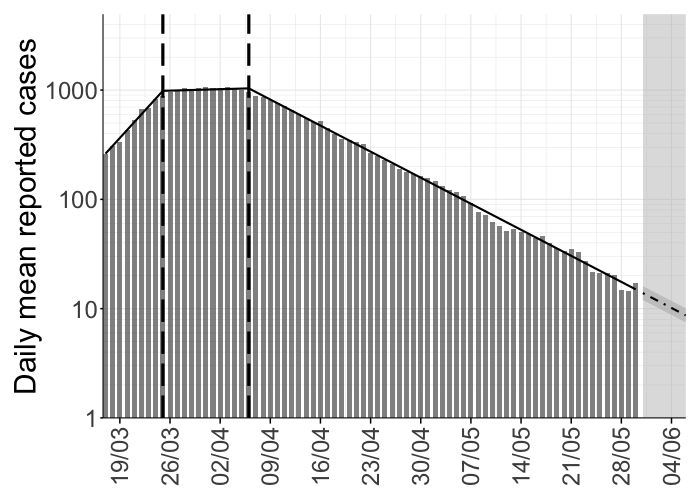}}
\subfigure[Germany.\label{}]{
\includegraphics[width=5cm]{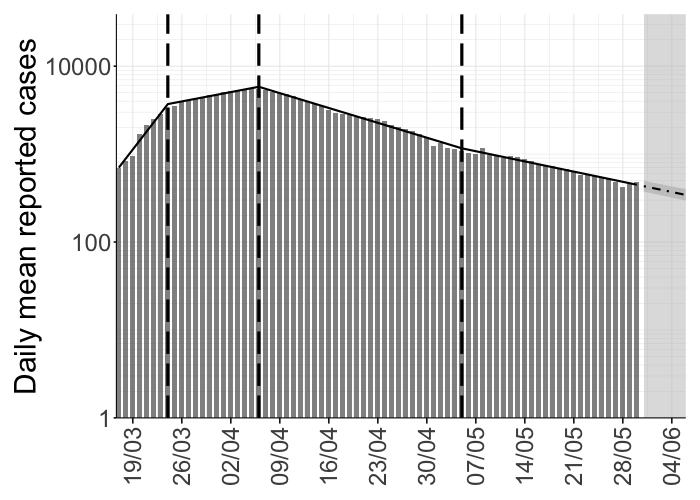}}
\subfigure[Italy.\label{}]{
\includegraphics[width=5cm]{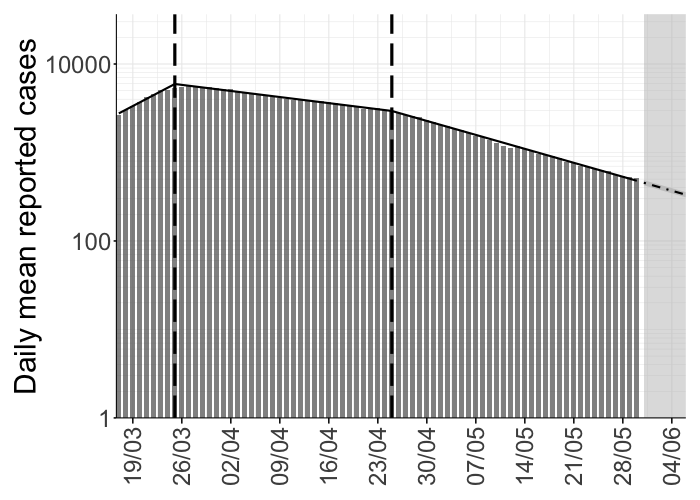}}
\caption{Linear splines curve fitting of mean daily new registered cases of Covid-19 in logarithmic scale for eight countries considering data until May 30th. In each graph, the gray bars indicate mean daily recorded new cases, the solid black line is the fitted curve, dotted black line is the prediction for the following 7 days, the shaded area indicates the interval of estimates, vertical dashed lines represent the knots locations.
}\label{fig:covidlinear}
\end{figure}

Considering data available until May 30th 2020, Figure~\ref{fig:covidcubico} shows the curve fitting using 
natural cubic splines to data in linear scale for Brazil, USA, Spain, South Korea, Iran, Switzerland, 
Germany, and Italy, eight countries that are in different stages of the epidemic. 
Although linear scale is better for visualization of peaks and/or oscillations and natural cubic splines are suitable for a smooth fitting, the change points are better determined when we consider linear splines as splines basis to the data transformed to logarithmic scale. For this reason, we also present curve fitting for the same eight countries using linear splines on data in logarithmic scale in Figure~\ref{fig:covidlinear}.  

In the European countries considered here, we note that mean daily new cases present a decreasing 
pattern as the peak occurred in late March, period in which our method selects some knots, 
indicating a change in trend. 
On the other hand, the daily new reported cases are in a high plateau in the USA, since the right-
hand side of the curve is nearly parallel to the horizontal axis of Figure \ref{fig:covidlinear}(b).
Moreover, the first change point in late March indicates a decrease in the slope of the fitted curve, 
the remaining knots are located near the date the country reaches its peak so far.
In Brazil, the data suggest the country have not reached its peak of infection yet, change points 
merely reveal small changes in the increasing pattern.
Finally, Iran and South Korea both show oscillations during the observed period, however at latest 
dates the new cases are apparently increasing, suggesting a new wave of reported cases.
We also perform similar analysis for several other countries and for Brazilian states
 and municipalities, that are updated daily, and the results and coded algorithms in  {\tt R} can be accessed at the webpage 
 \url{https://www.ime.usp.br/~gpeca/covid-19}.

We mention that the principal aim of this section is solely to show a descriptive analysis of the Covid-19 situation in the countries based on public available datasets and the proposed automatic method of knots estimation as a tool to do so. For a public policy guide to deal with this sad pandemic context, several social and biological factors beyond the scope of our study should be considered on a deeper analysis. 

\section{Discussion}
In this paper we introduced a method to estimate the number and position of the knots of
a spline regression function. The method is based on the minimization of the sum of squared
residuals plus a penalty that depends on the number of knots. We evaluated the performance of the
criterion on simulated data, and we showed that our proposed method had a great 
performance 
in the simulation studies, even with low SNR values. We also applied the method to perform a
 descriptive data analysis on Covid-19 daily reported cases in several countries.
In this analysis we showed that the penalized least squares estimation
guaranteed quite smooth curve fittings that estimated accurately underlying smooth 
functions automatically, i.e, without requiring to specify the number of internal knots and 
their locations, which is necessary in most spline based curve fitting methods available 
in the literature.

The penalizing constant $\lambda$ used in our data analyses was set to a fixed value. 
However as in many other similar approaches, it could be chosen by cross-validation procedure. 
From a theoretical point of view, one open question 
that remains is what is the rate for the penalizing constant $\lambda$ in order to
obtain consistency  of the estimator $\hat\theta$. 
Some results on this direction for related penalized models 
is presented in Castro et al. (2018) and Leonardi et al. (2021), where a consistency result was proved for a penalizing constant of order $n^{-1/2}$. 
As future work, we will study whether the consistency also holds in the context of our work.

\section*{Acknowledgements}

This article was produced as part of the activities of FAPESP's\footnote{S\~ao Paulo Research Foundation, Brazil}  Research, Innovation and Dissemination Center for Neuromathematics, grant 2013/07699-0, and FAPESP's project ``Model selection in high dimensions: theoretical properties and applications,  grant 2019/17734-3. A.S. is supported by a CAPES\footnote{Coordination of Superior Level Staff Improvement, Brazil} fellowship. M.T.F.S. is supported by a CNPq\footnote{National Council for Scientific and Technological Development, Brazil} fellowship. F.L is partially supported by a CNPq research fellowship, grant 311763/2020-0.

\end{document}